\documentclass[aps,floatfix,superscriptaddress,tightenlines,amsmath,amssymb,nofootinbib,10pt,longbibliography,twocolumn,prx]{revtex4-2}

\usepackage{algorithm}
\usepackage[noend]{algpseudocode} 
\usepackage{lineno}
\usepackage[utf8]{inputenc}
\usepackage{graphicx}
\usepackage{csquotes}
\usepackage{multirow} 
\usepackage{amsmath}
\usepackage[bb=pazo]{mathalpha}
\usepackage{mathrsfs} 
\usepackage{hyperref}
\usepackage{braket}
\usepackage{xcolor}
\usepackage{bbm}
\usepackage{comment}
\usepackage{enumitem} 
\usepackage{siunitx}
\usepackage[sort&compress]{natbib}
\usepackage{wrapfig}
\usepackage{array}
\newcolumntype{L}[1]{>{\raggedright\let\newline\\\arraybackslash\hspace{0pt}}m{#1}}
\newcolumntype{C}[1]{>{\centering\let\newline\\\arraybackslash\hspace{0pt}}m{#1}}
\newcolumntype{R}[1]{>{\raggedleft\let\newline\\\arraybackslash\hspace{0pt}}m{#1}}

\begin{document}

 \title{
Quantum limited displacement estimation with substantially enhanced dynamic range \\ using frequency-resolved two-photon interference  }

 \title{
Enhanced dynamic range for optical delay estimation by sampling the frequencies of two interfering photons}

 \title{
Quantum-limited optical delay sensing across an enhanced dynamic range by frequency-resolving two-photon interference}

\author{Russell M. J. Brooks}
\affiliation{Institute of Photonics and Quantum Sciences, School of Engineering and Physical Sciences, Heriot-Watt University, Edinburgh EH14 4AS, United Kingdom}
\author{Luca Maggio}
\affiliation{School of Mathematics and Physics, University of Portsmouth, Portsmouth PO1 3QL, United Kingdom}
\author{Thomas Jaeken}
\affiliation{Institute of Photonics and Quantum Sciences, School of Engineering and Physical Sciences, Heriot-Watt University, Edinburgh EH14 4AS, United Kingdom}
\author{Joseph~Ho}
\affiliation{Institute of Photonics and Quantum Sciences, School of Engineering and Physical Sciences, Heriot-Watt University, Edinburgh EH14 4AS, United Kingdom}
\author{Erik M. Gauger}
\affiliation{Institute of Photonics and Quantum Sciences, School of Engineering and Physical Sciences, Heriot-Watt University, Edinburgh EH14 4AS, United Kingdom}

\author{Vincenzo Tamma}
\affiliation{School of Mathematics and Physics, University of Portsmouth, Portsmouth PO1 3QL, United Kingdom}

\author{Alessandro Fedrizzi}
\affiliation{Institute of Photonics and Quantum Sciences, School of Engineering and Physical Sciences, Heriot-Watt University, Edinburgh EH14 4AS, United Kingdom}

\begin{abstract}
Optical sensing schemes that rely on two-photon interference provide a powerful platform for precision metrology, although they are inherently constrained by a trade-off between dynamic range and measurement precision. To overcome this limitation, we sample the frequencies of two interfering photons, which extends the sensitivity in the time domain. This enhances the dynamic range of optical delay estimation by up to twenty times compared to the non-resolved estimates. We demonstrate this approach with independent photon sources and show the behaviour of finite frequency resolving detectors. This technique enables scan-free nanometre resolution depth sensing over a millimetre-scale range, with applications in biological and nanomaterial imaging. 

\end{abstract}

\maketitle

Two-photon interference is a consequence of the quantum nature of the electromagnetic field, and is foundational to many quantum optics experiments, from photonic quantum computing and communication to metrology and sensing~\cite{hong1987measurement, rarity2020nonclassical, tamma2015multiboson, Pan1998, Knill2001, Browne2005,pirandola2018advances,Danilo23,laibacher2015physics,muratore2025superresolution,ho2024quantum,faleo2024entanglement}. Indistinguishable photons impinging on a balanced beam splitter manifest the non-classical tendency to exit through the same output port from the destructive interference of the two-photon probability amplitudes~\cite{hong1987measurement, rarity2020nonclassical,fearn1989theory, tamma2015multiboson}. This phenomenon has been exploited for precision estimation of optical path delays and coherence measurements, typically achieving sub-picosecond~\cite{steinberg1993measurement,branning2000simultaneous,dauler1999tests,steinberg1992dispersion} and recently attosecond temporal precision~\cite{lyons2018attosecond,chen2019hong,lualdi2025fast}. In comparison, direct time-of-flight using single-photon detectors is limited by detector jitter on the order of tens of picoseconds~\cite{pellegrini2000laser,rapp2021high}. However, methods that rely on measuring only the photon statistics at the output of a beam splitter are limited by the photons’ temporal bandwidth, which constrains the operational range and prevents sensitive measurements once the delay exceeds the coherence time~\cite{lyons2018attosecond, NoiseyHOM, scott2020beyond}.

Conventional multiphoton interference experiments do not resolve the photonic inner modes, such as frequency and time, which ignores the useful quantum information encoded within the photons~\cite{laibacher2015physics, laibacher2018symmetries}. It was shown theoretically in Ref.~\cite{laibacher2018symmetries} that resolving the photon frequency can restore interference even between non-identical photons and was later experimentally demonstrated between temporally distinguishable photons~\cite{orre2019interference}. Similarly, frequency-resolved two-photon interference reveals quantum beat notes in the interferogram~\cite{gerrits2015spectral, jin2015spectrally, jin2021NOON}, in contrast to the single fringe of the non-resolved interferogram~\cite{hong1987measurement}, accessing the spectral correlations that can be harnessed for precision metrology.

Recently, metrological schemes that sample the photonic inner modes have been shown to achieve the ultimate quantum-limited sensitivity in estimating conjugate variables, including time–frequency shifts~\cite{Danilo23, maggio2024freq}, transverse position–momentum displacements~\cite{triggiani2025momentum, triggiani2024estimation}, and full-state polarisation~\cite{maggio2025multi}. In optical delay estimation, ideal frequency-resolved sampling offers uniform sensitivity across the entire delay range, unlike the non-resolved method, which is parameter-dependent and must be optimised to reach maximum sensitivity~\cite{lyons2018attosecond}. However, these theoretical works require infinite resolving capabilities to fully erase information in the conjugate space and restore quantum interference~\cite{scully1982quantum, kim2000delayed}. In the absence of such devices, experiments are limited by finite resolution. The interference visibility depends on the spectral resolution, which must satisfy $\delta \omega \ll 1/\lvert \Delta t \rvert$ to restore interference with temporal delay $\Delta t$~\cite{Danilo23}. This has been studied using non-degenerate spectrally entangled photons~\cite{chen2022spectroscopy}, and weak coherent states~\cite{di2025high}.  

In this work, we report on the metrological advantage of frequency-resolved sampling for optical delay estimation with independent, spectrally uncorrelated single-photon states. We employ high-resolution ($6.9/2\pi$~GHz) frequency-resolved detectors, benchmark this method against a maximum-likelihood estimator without the spectral information and realise more than a twenty-fold increase in dynamic range and a thousand-fold improvement relative to other implementations~\cite {lyons2018attosecond,chen2019hong,lualdi2025fast}. We analyse the effects of finite frequency resolution on the Fisher information and discuss the practical utility of this technique for precision depth sensing.

\begin{figure}[htbp]
    \centering
    \includegraphics[width=1\linewidth]{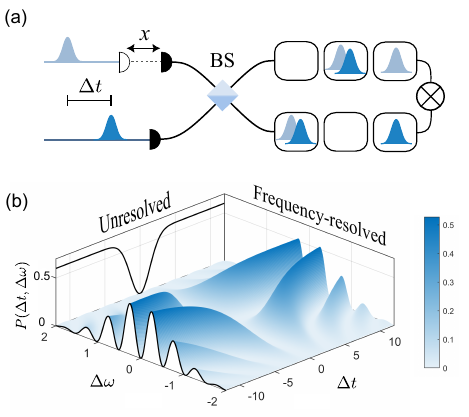}
    \caption{ (a) Two-photon interference setup. Two photons are input on a 50:50 beam splitter (BS) while a translation stage scans the path length delay $x$. Detectors placed on each exit port of the beam splitter can be either bucket detectors or frequency-resolving detectors. 
    (b) The two-photon frequency-resolved coincidence outcome probability, Eq.~\ref{FRout} is plotted as a function of the time delay $\Delta t$ and frequency difference $\Delta \omega$. Integrating over all frequencies results in the familiar HOM-dip along the non-resolved axis.}
    \label{fig:concept}
\end{figure}

We now describe the method for estimating the relative time delay; full derivations are provided in section~\ref{SM} of the Supplemental Material (SM). Consider two non-entangled photons arriving at separate ports of a balanced beam splitter at times $t_1$ and $t_2$, with a relative delay of $\Delta t=t_1-t_2$, as shown in Fig.~\ref{fig:concept}. Each photon has a Gaussian spectrum with a bandwidth given by the standard deviation $\sigma$ and a corresponding temporal bandwidth $\tau=1/2\sigma$. The distinguishability in all other degrees of freedom is characterised by the parameter $\eta\in[0,1]$, with $\eta=0$ representing complete orthogonality and $\eta=1$ denoting perfect overlap. When the photons arrive simultaneously at the beam splitter, they interfere destructively and ‘bunch’, exiting from the same output port with a probability determined by their temporal overlap. Temporal distinguishability leads to photons exiting different ports and subsequently being detected as a coincidence event. The photon-pair distribution after interference at the beam splitter is given by~\cite{lyons2018attosecond, scott2020beyond},
\begin{align}
    P_{\delta}^{\eta}(\Delta t) &= \frac{1}{2}\left(1+ \delta  \eta^2\mathrm{e}^{-\sigma^2\Delta t^2}\right), \label{NRout} 
\end{align} 
where $\delta = +1, -1$ are the bunching or coincidence outcomes, respectively, of a single measurement round.

When resolving the inner degree of freedom, namely frequency $\omega$, the photon-pair distribution exhibits interference fringes far beyond the coherence time~\cite{gerrits2015spectral, jin2015spectrally}, as described by~\cite{Danilo23, laibacher2018symmetries},
\begin{equation}
         P_{\delta}^{\eta}(\Delta t, \Delta \omega) = \frac{\mathcal{C}(\Delta \omega)}{2} \Bigl(1+ \delta \eta^{2} \cos(\Delta \omega \Delta t) \Bigr), \label{FRout}
\end{equation}
where $\Delta \omega =\omega_1-\omega_2$ is the difference in frequency and $\mathcal{C}(\Delta \omega)$ is the joint spectral distribution, see~\ref{supp_prob}. For a fixed delay $\Delta t$, the output distributions in Eqs.~\ref{NRout} and~\ref{FRout} can be sampled by measuring both the bunching ($N_B$) and coincidence ($N_C$) outcomes. In addition, sampling the joint frequencies $(\omega_1, \omega_2)$ is used to find the frequency difference $\Delta \omega$. The time delay $\Delta t$ is then inferred through maximum likelihood estimation (MLE), see~\ref{supp_prob}. 

The non-resolved estimation (NR) only requires the measurement of the total number of bunching ($N_B$) and coincidence ($N_C$) events~\cite{scott2020beyond, lyons2018attosecond}. The corresponding estimator can be expressed analytically:
\begin{align}
   \Delta t_{\text{NR}}(N_C, N_B) &= \frac{1}{\sigma} \sqrt{\log\left(\eta^2 \frac{N_B + N_C}{N_B - N_C}\right)}.\label{nrestim} 
\end{align}

In contrast, the frequency-resolved distribution given in Eq.~\ref{FRout} does not yield an analytic solution for $\Delta t$. Instead, we construct the estimator numerically via the likelihood function. Let the sample of $N$ measured outcomes be denoted by $S_N = \{s^{\{i\}}\}_{i=1}^N$, where each $s^{\{i\}} = (\Delta\omega^{\{i\}}, \delta^{\{i\}})$ contains a frequency difference and bunching/anti-bunching outcomes. The likelihood of a delay value $\Delta t$ given the sample is defined as $\mathcal{L}_R(\Delta t \mid S_N) = P(S_N \mid \Delta t)$. The estimator $\widetilde{\Delta t}_{\text{MLE}}$ is the value with the largest log-likelihood, see~\ref{SM_esti_algo},
\begin{equation}
    \mathcal{L}_R(\widetilde{\Delta t}_{\text{MLE}} \mid S_N) = \sup_{\Delta t} \mathcal{L}_R(\Delta t \mid S_N).\label{res_estim}
\end{equation}
The only unknown parameter remaining in Eq.~\eqref{FRout} is $\eta$, which can be obtained experimentally in the calibration step. Therefore, the frequency-resolved scheme provides a numerical estimate of the time delay by matching the measured outcomes to the theoretical distribution via maximum likelihood. The error on these parameters will feed forward into the variance of the estimator which we discuss in~\ref{SM_nosie}. A full derivation of the estimator in Eq.~\eqref{res_estim} is provided in Ref.~\cite{Danilo23}.

We perform the experiment using two configurations. The first uses heralded single photons from two independent parametric down-conversion (PDC) sources, each with Gaussian spectra and without spectral correlations. The second configuration uses a single PDC source with higher pair rates and indistinguishability, see~\ref{SMsetup}. Each photon-pair source consists of a domain-engineered, aperiodically poled KTP crystal~\cite{pickston2021optimised, Graffitti:18}, designed for group-velocity matching with a pulsed pump laser. This enables the generation of spectrally pure, mutually indistinguishable photon pairs at \SI{1550}{\nano\metre}~\cite{Francesco_crystal, Graffitti_2017}, see supplement for rates~\ref{SM}. 

\begin{figure}
    \centering
    \includegraphics[width=1\linewidth]{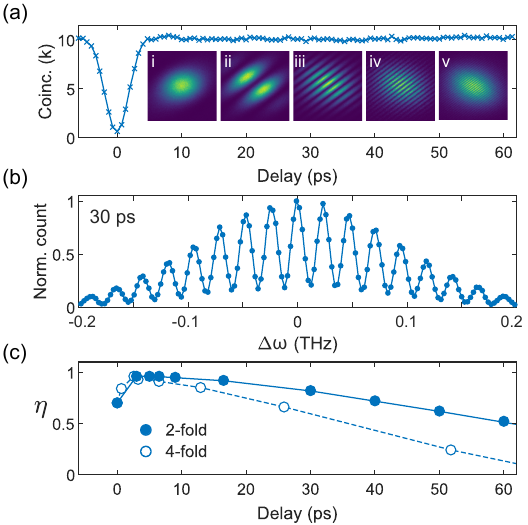}
    \caption{(a) Two-photon interference over the entire delay range, measuring coincidence counts from a single source (2-fold) over a 20~mm range and converted to units of time in picoseconds. Insets (i-v) show the joint spectral intensity (JSI) for $[0, 6, 16, 30,60]$ ps delays. The (2-fold) marginal for the JSI at 30~ps is presented in (b). The fringe visibility is found to be 0.78, which is equivalent to $\eta$. Indistinguishability $\eta$ as a function of delay is plotted in (c), extracted by fitting the marginals to the theoretical distribution in Eq.~\ref{FRout}, for both the single source (2-fold) and independent sources (4-fold). The solid and dashed lines are fitted using a spline and follow a linear trend after the maximum $\eta$ value.}
    \label{fig:JSI_HOM}
\end{figure}

We prepare the photons in identical quantum states by matching their frequency, polarisation, and spatial mode, leaving only the relative time delay $\Delta t$ at the beamsplitter to be varied using a mechanical translation stage that is calibrated using standard HOM interference Fig.~\ref{fig:JSI_HOM}(a).

To resolve the photon frequencies we use a time-of-flight (ToF) spectrometer  comprising a \SI{50}{\kilo\metre} fibre spool and a picosecond-resolution time-tagger, see supplement~\ref {SM}. This setup achieves a spectral resolution $\delta \omega$ of $6.9/2\pi$~GHz, which sets the maximum bound on the delay range as $\Delta t \ll 144$~ps using the relation $\delta \omega \ll \frac{1}{|\Delta t|}$. The joint spectral intensity (JSI) for the bunching outcomes using a single source is plotted in Fig.~\ref{fig:JSI_HOM}(a). The marginal of the JSI reveals the interference visibility as a function of $\Delta t$. The visibility is related to the indistinguishability $\eta$ through Eq.~\ref{FRout} which we find by fitting $\eta$ to the distribution, see Fig.~\ref{fig:JSI_HOM}(c). We see that for larger delays, the photon indistinguishability goes down because a finer frequency resolution is required to erase all the distinguishable information between the photons. Furthermore, the 4-fold indistinguishability is lower due to additional noise from the extra detectors and fibres.

\begin{figure}
    \centering
    \includegraphics[width=1\linewidth]{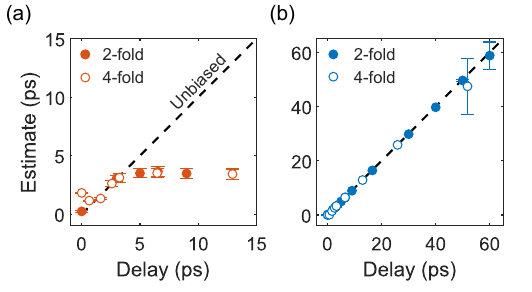}
    \caption{Accuracy of estimated delays using (a) non-resolved (NR) methods and (b) the frequency resolved (FR) method with a single source (filled) or independent source (not filled). Two-fold estimates use $r = 1000$ repeats, each using $ N = 1000$ samples, while the four-fold points use $r=100$ estimates with $N = 100$ samples. Error bars are calculated from one standard deviation of the set of estimates $\{\widetilde{\Delta t}_i \}^r_{i=1}$ rounds.}  
    \label{fig:accuracy}
\end{figure}

\begin{figure}
    \centering
    \includegraphics[width=1\linewidth]{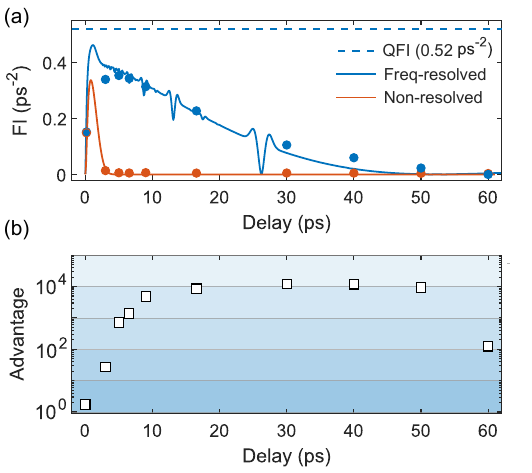}
    \caption{(a) The experimental metrological information $\mathscr{F}_{exp}$ for the 2-fold estimates with frequency-resolved (FR) and non-resolved (NR) measurements is plotted against the prepared delay. The non-resolved Fisher information is plotted as a solid orange line using Eq.~\eqref{NR_FI} and $\eta=0.98$, displaying a peak at around 1 coherence length, while the frequency-resolved Fisher information is plotted as a solid blue line using Eq.\ref{finiteFI} and displays an extended sensitivity range. The blue dashed horizontal line is the QFI with value 0.52~($\text{ps}^{-2}$). (b) The metrological advantage is plotted for each estimate in logarithmic scale using the ratio of the resolved and non-resolved mean squared error, $\mathrm{MSE_{NR}/MSE_{FR}}$.}
    \label{fig:precision}
\end{figure}

We evaluate the performance of the frequency-resolved and non-resolved estimation strategies for both source configurations by comparing their accuracy and precision. Figure~\ref{fig:accuracy} shows the estimated delay plotted against the prepared delay for both methods. Each data point represents the average over $r$ estimates, each consisting of $N$ samples (see figure caption for details). The non-resolved estimates remain unbiased up to 3~ps, then the estimator begins to fail. In comparison, the averaged frequency-resolved estimate is relatively unbiased up to 60~ps, but with a larger variance. The ratio of the mean squared errors (MSE) is plotted in the inset to Fig.~\ref{fig:precision}(a); this ratio demonstrates the improved uncertainty of the frequency-resolved estimates over the non-resolved estimate by up to four orders of magnitude. 

We quantify the theoretical time-delay precision using Fisher information analysis.  In the presence of both partial indistinguishability $\eta$ and loss $\gamma$, the Fisher information of the frequency-resolved estimate is given by~\cite{Danilo23}
\begin{equation}
\begin{aligned}
    \mathscr{F} &= \mathbb{E}\Big[ \Big( \frac{\text{d}}{\text{d} \Delta t} \log P_{\delta}^{\eta}(\Delta t, \Delta \omega) \Big)^2 \Big] \\
    &= \eta^4 \gamma^2 \int_{\mathbb{R}} \text{d}\Delta \omega ~\mathcal{C}(\Delta \omega)^2 \frac{\sin^2(\Delta \omega \Delta t)}{1-\eta^4 \cos^2(\Delta \omega \Delta t)} \\
    &= \eta^4 \gamma^2 \mathcal{I}(\Delta t), 
\end{aligned} \label{FI_FR}
\end{equation}  
where $\mathcal{I}(\Delta t)$ is an integral over the joint spectral distribution. When using post-selected measurements, we can rescale the total number of resources and thus drop the $\gamma$ term. The maximum information that can be attained over all possible measurement strategies is called the quantum Fisher information (QFI). For non-entangled photons with Gaussian spectra, the QFI is related to the spectral bandwidth by $\mathscr{H} = 2\sigma^2 = 1/2\tau^2$, 
which is half as much as for entangled photons~\cite{scott2020beyond, chen2019hong}.

It was shown in Ref.~\cite{Danilo23} that the frequency-resolved Fisher information with Gaussian spectra and perfect indistinguishability reaches the quantum limit, $\mathscr{F}_{\eta=1}(\Delta t) = \mathscr{H}$. This demonstrates that when photons differ only in arrival time, the frequency-resolving scheme is optimal for extracting the most information from the photons. For $\eta < 1$, the inequality \( \mathscr{F}_\eta(\Delta t) \geq \mathscr{F}^{NR}_\eta(\Delta t) \) holds due to the convexity of the Fisher information and the marginalisation over spectral modes. By contrast, the non-resolving protocol Fisher information is given by~\cite{scott2020beyond, lyons2018attosecond},
\begin{equation}
    \mathscr{F}^{\text{NR}}_\eta(\Delta t) = \frac{\eta^4 \mathscr{H}}{\exp\left(\frac{\Delta t^2}{2\tau^2}\right) - \eta^4} \frac{\Delta t^2}{2\tau^2}. \label{NR_FI}
\end{equation}
This expression shows that the sensitivity rapidly decays for delays $\Delta t \gg \tau$, as the interference vanishes outside the coherence window.

So far, this analysis assumes infinite frequency resolution. To account for realistic detection, we model the finite spectral resolution $\delta \omega$ by dividing the frequency axis into discrete bins of width $2\delta \omega$. Each bin \( n \) has an associated probability$ P_{n,\delta}^{\eta}(\Delta t, \delta \omega, \sigma) =\mathcal{P}_n $, where $ \delta \in {+1, -1}$ denotes the output port. To obtain $( \mathcal{P}_n )$, we convolve the ideal joint spectrum with a triangular detector response function, reflecting the finite frequency resolution. This model captures how frequency resolution degrades sensitivity, offering a practical framework to evaluate performance under experimental conditions. A full derivation is provided in the Supplementary Material~\ref{SM_finite_freq}. The total Fisher information with finite frequency resolution is then, 
\begin{equation}
  \mathscr{F}(\Delta t)=\sum_{n=0}^{n_{\mathrm{MAX}}}\mathscr{F}_n(\Delta t).  \label{finiteFI}
\end{equation}

The precision of the experimental estimates against the theoretical bound is calculated from the experimental metrological information, $\mathscr{F}_{exp}$, using the variance of the estimates via the Cramér–Rao bound~\cite{feller1947harald,rohatgi2015introduction} (CRB), $\mathrm{var}[\widetilde{\Delta t}] \geq 1/N \mathscr{F}_{exp}$,
where $N$ is the number of samples per estimate. Fig.~\ref{fig:precision} presents both the experimentally obtained metrological information and the theoretical Fisher information across the full delay range for the non-resolved (a) and frequency-resolved (b) estimates. The Fisher information of the non-resolved estimate (solid line) is calculated using Eq.~\ref{NR_FI}, with the maximum photon indistinguishability $\eta = 0.98$ obtained from calibration, see~\ref{SMsetup}. The Fisher information peaks at 0.336~$\text{ps}^2$ and falls off with a width approximately $2\tau \approx 2$~ps. The experimental data align well with the theoretical prediction, confirming the loss of precision at delays greater than $2\tau$.

In frequency-resolved metrological information is plotted in Fig.~\ref{fig:precision}(a) using Eq.~\ref{finiteFI}, where both $\eta$ and $\Delta t$ vary, and the values of $\eta$ are found from Fig.~\ref{fig:JSI_HOM}(c). The experimental values agree closely with theory for the 2-fold configuration. However, the 4-fold estimates achieve a metrological lower Fisher information than expected. This reduced performance may result from the increased complexity of the heralded setup, including additional noise from fibre paths and the cumulative effect of timing jitter across twice as many detectors.

Photon indistinguishability plays a critical role in both estimation strategies, as $\mathscr{F} \propto \eta^4$. To maximise the Fisher information and approach the QFI, we reduced the pump power of the single-source setup to 100~mW, thereby increasing the two-photon interference visibility, with fitted indistinguishability $\eta = 0.98$. At this level, the experimentally obtained metrological information reaches $\mathscr{F} = 0.442 ~(\text{ps}^{-2})$, which corresponds to 0.85 of the QFI.

We test the precision limits of our experimental setup by resolving a 3~fs shift between two delay positions. We picked a reference delay 6.567~ps ($\approx 3\tau$) and estimated this delay using 7.9 million samples at 100~mW pump power. We then translated the delay stage by 1~$\mu$m, which corresponds to a 3~fs shift, the smallest displacement the translation stage can perform with a reliability of 0.5~$\mu$m. We used the same number of rounds to estimate the new delay with uncertainty of 0.5~fs, see~\ref{SM_micro_shift}. We measured a shift of 4~fs, with an error of 1~fs, the discrepancy most likely coming from the accuracy of the delay stage. The chosen delay, 6.567~ps, lies in the region where the non-resolved strategy fails to provide unbiased estimates, further demonstrating the sensitivity enhancement from frequency resolving. 

Resolving the frequency difference between interfering photons greatly enhances the dynamic range of optical delay estimation relative to the identical non-resolved method. The achievable precision depends on both the spectral bandwidth and the resolution of the frequency-resolving detectors, which can be improved in future experiments. We derived the Fisher information for finite frequency resolution in~\ref{SM_finite_freq}. Improving the resolution by a factor of ten extends the dynamic range also by a factor of ten, and an increase by a factor of sixty nearly saturates the quantum Cramér-Rao bound over the full 60~ps delay range. Compared with a recent experiment that reached attosecond precision ~\cite{lyons2018attosecond}, we obtained a similar per-sample performance even after incorporating their loss model. The best measured precision in Ref.~\cite{lyons2018attosecond} was 4.7 attoseconds using 414 billion samples, while we reached 500 attoseconds using 7.9 million samples. The usual trade-off between precision and dynamic range appears to be avoided for delays up to 10~ps with the frequency-resolved method. In this regime, the Fisher information exceeds the peak of the non-resolved Fisher information, so the need for careful optimisation near that peak is removed while retaining equal or greater precision.

The main limitation of our present setup is loss from chromatic dispersion using standard single-mode fibres for time-of-flight spectrometry, which is about 99\% for 50 km spools. This can be reduced by using commercial single-photon spectrometers with lower loss, a chirped fibre Bragg grating \cite{orre2019interference}, or a diffraction grating combined with a SPAD array~\cite{chen2023spectrally, di2025high}. However, the resolution from these devices is typically below that of time-of-flight spectroscopy with dispersive fibres. The precision of our setup can be further enhanced using entangled photon pairs to realise a $\sqrt{2}$ improvement per sample~\cite{chen2023spectrally}. Further gains are possible with spectrally engineered sources that use detuned bi-photon schemes \cite{lualdi2025fast,chen2019hong}, which can provide large Fisher information without fringe tracking or calibration, overcoming the sensitivity trade-off in optical sensors~\cite{bobroff1993recent,fraden2004handbook}.

Another advantage of two-photon interference is the robustness to phase noise and imbalanced losses, which significantly impact the measurement precision of phase-sensitive interferometers~\cite{NoiseyHOM}. The sensitivity of our approach is also comparable to phase-based methods but without active stabilisation~\cite{rarity2020nonclassical, fearn1989theory} or the phase wrapping problem, which requires additional phase tracking algorithms~\cite{Itoh:82, Hayashi_2019}.

In conclusion, we have demonstrated a high-precision and wide dynamic range optical delay sensor using frequency-resolved two-photon interference. This technique is scan-free and well-suited for integrated photonic and nanoscale sensing, for example, characterising substrate layers or aligning micro-scale components where large working distances and nanometre-scale precision must be maintained without recalibration. The same approach can improve time transfer protocols by delivering sub-femtosecond precision while tolerating drifts between channels on the order of tens of picoseconds. 

\vspace{1em}
\noindent
\textbf{Acknowledgements}
\newline
We thank A. Pickston for helping characterise the photon source.
This work was supported by the UK Engineering and Physical Sciences Research Council (Grant Nos. EP/T001011/1.).
L. Maggio acknowledges partial support
by Xairos Systems Inc. 
V. Tamma acknowledges partial support from
the Air Force office of Scientific Research under award
number FA8655-23-1- 7046.

\vspace{1em}
\noindent
\textbf{Author contributions}
\newline
RB, AF and VT conceived the project. RB, TJ and JH performed the experiment, RB, TJ and LM analysed the results.
LM and VT developed the theoretical tools used in the analysis.
All authors contributed to writing and revisions of the manuscript.

\vspace{1em}
\noindent
\textbf{Competing interests}
\newline
The authors declare no competing financial or non-financial interests.

\bibliography{bib}
\onecolumngrid

\setcounter{equation}{0}
\setcounter{figure}{0}
\makeatletter

\setcounter{secnumdepth}{2}

\renewcommand{\thefigure}{SM\arabic{figure}}
\renewcommand{\theequation}{SM\arabic{equation}} 
\renewcommand{\thesection}{SM\arabic{section}}  

\begin{center}
    \textbf{\large \label{SM}Supplemental Materials}
\end{center}

\section{Experimental Setup}\label{SMsetup}

\begin{figure}
    \centering
    \includegraphics[width=0.7\linewidth]{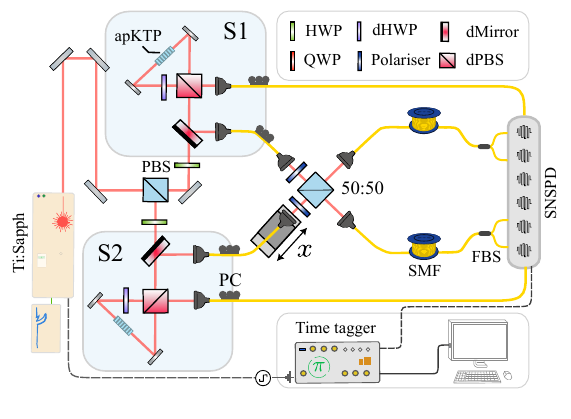}
    \caption{\textbf{Experimental layout with two sources}. A 775~nm pulsed Ti:Sapphire laser with 1.3~ps duration and 80~MHz repetition rate pumps two photon pair sources. The sources comprise of a  aperiodicity-poled KTP crystal housed in a Sagnac configuration, generating spectrally degenerate Gaussian profile photon pairs at 1550~nm. One photon from each pair is separated using dichroic polarising beam splitters (dPBS) and coupled to a single-mode fibre (SMF) to be detected imediately on superconducting nanowire detectors (SNSPD) and time-stamped on Hydra harp time tagger. The other pair from each source is then coupled to SMF and launched into free space onto a 50:50 beam splitter. The polarisation of each independent photon is prepared using a linear plate polariser set to Horizontal to match the polarisation degree of freedom. The Spectral component is also matched by calibrating the temperature of the crystal ovens to yield the highest independent-HOM visibility, found to be 91\%. Each path exiting the beam splitter is coupled to 50km SMF to induce chromatic dispersion and delay the photon packets in time. A time-of-flight spectrometer converts the time difference between the arrival of a photon and the arrival of the Ti:Sapphire sync pulse to estimate the frequency component after experience dispersion. Finally each path is multiplexed into two spatial modes to act as a pseudo number-resolving detectors. }
    \label{fig:2source_setup}
\end{figure}

A detailed experimental layout of the independent source configuration is shown in Fig.~\ref{fig:2source_setup}. The experiment employs two configurations: (I) heralded photons from independent sources, and (II) a single photon-pair source. In configuration (I), two independent photon-pair sources (S1 and S2) are pumped by a Ti:sapphire laser. Each source generates a photon pair, with one photon from each pair separated using a polarising beam splitter (PBS) and coupled into a single-mode fibre (SMF), where it serves as a herald. Long-pass silicon windows are used to filter out the pump light. The remaining photons are also coupled into SMFs and launched onto a balanced 50:50 beam splitter (BS). Each photon input on the BS is prepared with horizontal polarisation using a linear polariser and in-fibre polarisation controllers. The output ports of the BS are each connected to a 50~km SMF spool, forming a time-of-flight spectrometer. These outputs are then split again into two fibres, forming pseudo-number-resolving detectors. Photons are detected using superconducting nanowire single-photon detectors (SNSPDs) with approximately 80\% quantum efficiency and 50~ps timing jitter.

Our sources generate 2000 pairs/mW when pumped with a pulsed Ti:Sapphire laser at a 80~MHz clock-rate and 1.3~ps pulse duration. After state preparation and measurement, we incur heavy losses. Using 450~mW pump power, we measure on average 434 pairs per second and 4 heralded pairs (4-folds) per second. These low rates are due to the 10~db loss of each time-of-flight spectrometer (20~db total), and a further one-half loss in the pseudo-number resolving detector due to twice the double bunch events. The high-resolution data set used 100~mW pump power, generating on average 113 pairs per second.

Measurement outcomes are categorised as either bunching or coincidence events. For each two-fold detection event, we record the outcome as $s^{i} = (\Delta \omega^i, \delta^i)$, where $\Delta \omega^i$ is the frequency difference and $\delta^i$ indicates the type of event (bunched or anti-bunched).

\section{Frequency-resolving detector}

The time-of-flight (ToF) spectrometer is made up of 50~km fibre spool with a total group delay dispersion (GDD) $\phi''= \SI{909}{\pico\second^2}$, combined with detector jitter, resulting in a spectral resolution $\delta \omega$ of $6.9/2\pi$~GHz. The frequency is given by $(\omega^i - \Omega) = (t^i - t_{\Omega})/\phi'' $, where $t^i$ is the arrival time of the $i^{th}$ photon triggered by a sync signal from our Ti:sapphire laser and $\Omega= \SI{193}{\tera\hertz}$ is the central frequency of our 1550~nm photons. Since only the frequency difference $\Delta \omega^i$ is used in the MLE, we can set $t_\Omega = 0$. The frequency of both output ports is then resolved as a linear function of the ``time-of-flight" $\Delta \omega^i = \Delta t^i/\phi''$ through fibre with a dispersion factor $1/\phi''=0.0011(4)$~nm/ps.

\section{Calibration and experimental parameters}
 
The essential photonic parameters are experimentally obtained and presented in Table~\ref{Table_params}.
The photon coherence length and indistinguishability are found from the two-photon interferogram using standard Hong-Ou-Mandel interference in the coincidence basis. The coherence length is given by the width of the HOM dip~\cite{branczyk2017hong} and is calculated using a Gaussian fit. The photon indistinguishability is well approximated by the fringe visibility; we measure a visibility of 0.98 using a single source configuration at 100~mW power.

The spectral bandwidth is approximately $81/2\pi$~GHz, given by the standard deviation of the frequency distribution $\sigma$. We measure the frequency distribution with our time-of-flight spectrometer while preparing no delay between the photons input on and we sample from the bunching outcomes. The temporal bandwidth is calculated from the spectral bandwidth using the relation $\sigma=1/2\tau$.

Time delays are calibrated using an non-resolved second-order interferogram as shown in Fig.~\ref{fig:JSI_HOM}(a). The centre of the two-photon interference fringe defines the nominal zero point for the relative delay $\Delta t$. The photons are temporally synchronised using a motorised linear translation stage, which adjusts the path length of one arm of the interferometer with micron precision (equivalent to \SI{3}{\femto\second} temporal resolution).

\begin{table}[t]
\centering
\small
\caption{Experimental parameters}
\label{tab:metrics}
\begin{tabular}{lcc}
\hline
Metric & Gaussian $\sigma$ & FWHM \\
\hline
Coherence length $\tau_{c}$ & 1.24 & 2.92 \\
Temporal bandwidth $\tau$ [ps] & 0.98 & 2.31 \\
Spectral bandwidth $\sigma/2\pi$ [GHz] & 81 & 190 \\
\hline
\end{tabular}\label{Table_params}
\end{table} 

\section{Estimation Protocol} \label{SM_esti_algo}

Our goal is to estimate the relative delay $\Delta t$ from the recorded measurement outcomes. The frequency-resolved distribution given in Eq.~\ref{FRout} lacks an analytic inverse for $\Delta t$, so we employ a maximum-likelihood estimation (MLE) procedure. The dataset consists of $N$ independent measurement outcomes, denoted as $S_N = \{s^{(i)}\}_{i=1}^{N}$, where each outcome is $s^{(i)} = (\Delta \omega^{(i)}, \delta_i)$, and $\delta_i$ denotes the detection outcome (bunching or coincidence). The likelihood function was introduced in Eq.~\ref{res_estim},
which gives the probability of observing the sample $S_N$ given a particular value of $\Delta t$. The MLE estimator $\widetilde{\Delta t}_{\text{MLE}}$ is obtained numerically by maximising the likelihood as,
\begin{equation}
    \widetilde{\Delta t}_{\text{MLE}} = \arg \max_{\Delta t} \, \mathcal{L}_R(\Delta t \mid S_N).
\end{equation}
This estimation protocol directly links the measured frequency-resolved coincidences to the temporal delay $\Delta t$ prepared in the experimental setup, and its sensitivity is governed by the indistinguishability parameter $\eta$ obtained during calibration. The estimation protocol is presented in Algorithm~\ref{SM_table1}.

\begin{algorithm}[H]
\caption{Maximum Likelihood estimation for frequency-resolved and non-resolved samples}
\label{SM_table1}
\begin{algorithmic}[1] 
\State \textbf{Input:} mode flag \texttt{resolved} $\in\{\texttt{true},\texttt{false}\}$, dataset
$\mathcal{D}$, visibility $\nu$, spectral width $\sigma$, grid of delays $\mathcal{T}=\{t_0,\ldots,t_{K}\}$, sample size $N$, repeats $R$
\State \textbf{Output:} estimates $\{\hat t_r\}_{r=1}^{R}$
\State If \texttt{resolved} is true then $\mathcal{D}=\{(\Delta\omega_n, x_n)\}_{n=1}^{M}$ with $x_n\in\{+1,-1\}$ else $\mathcal{D}=\{x_n\}_{n=1}^{M}$
\State Define single event probability for use when \texttt{resolved} is true
\For{$r=1$ to $R$}
  \State Draw $N$ indices without replacement $I_r\subset\{1,\ldots,M\}$
  \If{\texttt{resolved}}
    \State Initialise log likelihood vector $L(t)\gets 0$ for all $t\in\mathcal{T}$
    \For{each $n\in I_r$}
      \For{each $t\in\mathcal{T}$}
        \State $L(t)\gets L(t)+\log p(x_n,\nu,\Delta\omega_n,t,\sigma)$
      \EndFor
    \EndFor
    \State $\hat t_r\gets \arg\max_{t\in\mathcal{T}} L(t)$
  \Else
    \State $b\gets \#\{n\in I_r\,:\, x_n=+1\}$ \Comment bunching count
    \State $c\gets \#\{n\in I_r\,:\, x_n=-1\}$ \Comment coincidence count
    \If{$b\le c$ \textbf{or} $b=0$}
       \State \textbf{continue} to next repeat
    \EndIf
    \State $q\gets \dfrac{b+c}{\,b-c\,}$
    \If{$q\le 0$}
       \State \textbf{continue} to next repeat
    \EndIf
    \State $\hat t_r\gets \dfrac{\sqrt{\ln\!\bigl(q\,\nu^{2}\bigr)}}{\sigma}$
  \EndIf
\EndFor
\State \textbf{return} $\{\hat t_r\}_{r=1}^{R}$
\end{algorithmic}
\end{algorithm}  

\section{Estimator noise}\label{SM_nosie}
Let the measured frequency shifts be \(\{\Delta\omega_i\}_{i=1}^{N}\) with true density \(P_n(\Delta\omega)\).
We estimate \(\Delta t\) by maximising the log likelihood
$\ell(\Delta t) = \sum_{i=1}^{N} \log P(\Delta\omega_i \mid \Delta t)$,
which is equivalent to using \(P(\Delta t \mid \Delta\omega_i)\) up to a normalising constant. Assume a small model mismatch between the true density \(P_n\) and the model \(P(\,\cdot\,\mid \Delta t)\).
Write \(\Delta\omega_i = \tilde{\Delta\omega}_i + \varepsilon_i\), where \(\tilde{\Delta\omega}_i \sim P(\,\cdot\,\mid \Delta t)\) and \(\varepsilon_i\) captures the difference between the two distributions with
\(\mathbb{E}[\varepsilon_i]=0\) and \(\mathbb{E}[\varepsilon_i^{2}] = \sigma_{\varepsilon}^{2}\).
A second-order Taylor expansion of each log likelihood term about \(\tilde{\Delta\omega}_i\) gives,
\begin{equation}
\log P(\Delta\omega_i \mid \Delta t)
= \log P(\tilde{\Delta\omega}_i \mid \Delta t)
+ \partial_{\Delta\omega}\log P(\tilde{\Delta\omega}_i \mid \Delta t)\,\varepsilon_i
+ \tfrac{1}{2}\,\partial_{\Delta\omega}^{2}\log P(\tilde{\Delta\omega}_i \mid \Delta t)\,\varepsilon_i^{2}
+ o(\varepsilon_i^{2}).
\end{equation}
Taking expectations over the mismatch, the first order term vanishes since \(\mathbb{E}[\varepsilon_i]=0\).
The second order term contributes a bias proportional to \(\mathbb{E}[\varepsilon_i^{2}]\), hence to \(\sigma_{\varepsilon}^{2}\).
This contribution increases as \(P_n\) and \(P(\,\cdot\,\mid \Delta t)\) diverge. In the regime \(|\Delta\omega\,\Delta t|\ll 1\), the two distributions are practically equivalent, so \(\sigma_{\varepsilon}^{2}\) is negligible and the induced bias can be ignored to an excellent approximation.

\begin{figure}
    \centering
    \includegraphics[width=0.5\linewidth]{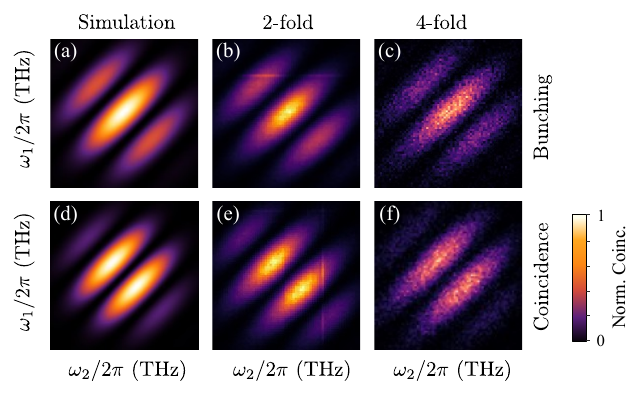}
    \caption{The joint spectral intensity (JSI) of two-photon interference with a 6.5ps delay. (a,d) are the theoretical simulation for the bunching and coincidence outcomes respectively. (b,e) are the dependent source JSI. (c,f) are the independent source JSI. The agreement between the single source and independent source JSI correlations confirms the apKTP Gaussian source can be used as an independent source of spectrally Gaussian distributed photons.   }
    \label{SM_JSI}
\end{figure}

\section{Spectral indistinguishability between one and two sources}

In this section, we show that using photons from one of our Gaussian sources is indistinguishable from two separate sources. This is crucial to ensure no spectral correlation exists between the two interfering photons, which may cause interference or mask the effect of the frequency-resolving contribution. We prepare both the single-source and two-source configurations with approximately 6.5~ps temporal delay and measure the joint spectral intensity from both bunching and coincidence outcomes, displayed in Fig.~\ref{SM_JSI}. We see excellent agreement between the simulated JSI, the single-source (2-fold) and the two-source (4-fold) interferogram. 

\section{Stability of the setup over time}

\begin{figure}
    \centering
    \includegraphics[width=1\linewidth]{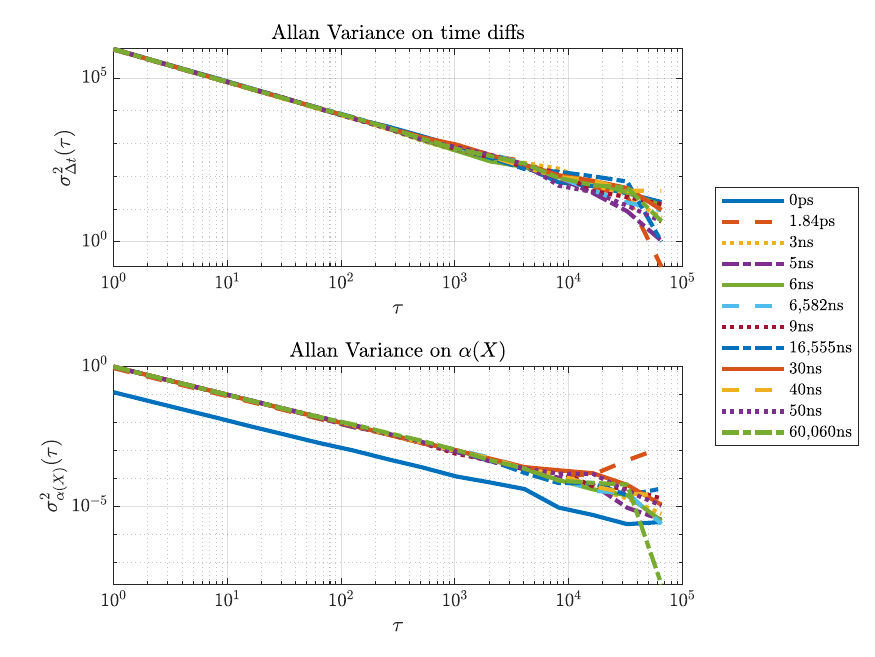}
    \caption{The Allan variance of the 2-fold samples is plotted to assess the stability of the estimates over a long period of time.} \label{Fig:Allan}
\end{figure}

Collecting many samples over a long period of time can cause bias in the estimates due to experimental drift. One way to mitigate this is to analyse the Allan variance of the experimentally obtained samples for the bunching and coincidence outcomes, $\alpha(X)$, where $X=1,-1$ and the spectral differences. In Fig.~\ref{Fig:Allan} we have plotted the Allan variance for the BoC and spectral differences, both the calculated using the following form,

\begin{equation}
\sigma_{y}^{2}(\tau)
=\frac{1}{2\,(M-1)}\sum_{j=1}^{M-1}\bigl(\tilde y_{j+1}-\tilde y_j\bigr)^2.
\end{equation}

Where $M=\left\lfloor \frac{N}{m} \right\rfloor$ is the disjoint clusters of length $m$ from $N$ samples. $\tilde y_j=\frac{1}{m}\sum_{i=1}^{m} y_{(j-1)m+i},
\qquad (j=1,2,\ldots,M)$.

\section{Maximum resolution estimation} \label{SM_micro_shift}

\begin{figure}
    \centering
    \includegraphics[width=0.7\linewidth]{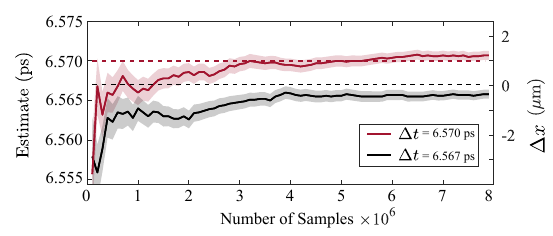}
    \caption{\textbf{Resolving micron shift} (a) The estimate bias has been plotted against the sample size for the independent source (4-fold) when preparing a $\Delta t=$6.54ps delay. Each data point is the average of 770 independent estimates and the error bars are calculated from the variance of the estimates scaled by $\Delta t$. The variance for the independent (FRI) and dependent (FRD) frequency-resolved estimates photon sources has been plotted on the second axis. The inset plot is the ratio of the QCRB and the variance of the FRD, on average we achieve 0.8 of the QCRB (b) Two delays are prepared using the minimum spatial step size of our linear stage, which is 1$\mu$m, or 3fs. The cumulative estimate has been plotted up to 7.9$\times 10^6$ samples. The error bars have been calculated from the CRB using the estimated $F_{\eta}^{G}=0.4ps^2$ per sample, this equates to a standard deviation of 562 attoseconds and a measurement precision of $\pm$168nm.}
    \label{SMmicron_shift}
\end{figure}

In this section, we analyse the resolution of the resolved optical delay estimation using our best prepared state. We use low pump power (100~mW) and achieve HOM visibility of 0.98, which improves the estimate precision. Fig.~\ref{SMmicron_shift} displays the cumulative estimate for the delays prepared at 6.567~ps and 6.570~ps.

\section{Derivation of Probability function in terms of frequency difference
}\label{supp_prob}
We start with eq.~3 from Ref. \cite{Danilo23}:
\begin{equation}
    P_{\eta}(\omega, \omega', X) = \gamma^2 \xi(\omega)^2 \xi(\omega')^2 
    \frac{ 1 + \alpha(X) \eta^2 \cos\left((\omega - \omega') \Delta t\right) }{2},
\end{equation}
where $\eta$ is the indistinguishably of the two photons, $\gamma$ is the detector efficiency, $X = B,C$ is the Bunching or Coincidence outcome of single measurement, $\alpha(B) = 1 $ and $\alpha(C) = -1$, $\xi(\omega)$ is the spectral distributions, that is assumed to be Gaussian, with a mean value $\Omega_0$ and a variance $\sigma^2$ for both the photons,
\begin{equation}
    \xi(\omega)^2=\frac{1}{\sqrt{2\pi\sigma^2}}\exp\left[-\frac{\left(\omega-\Omega_0\right)^2}{2\sigma^2}\right].\label{gauss}
\end{equation}
Since we notice that the quantum beats depends only to the difference of the detected frequencies, that is, $\Delta\omega=\omega'-\omega$, and not to their mean, we rewrite the probabilities applying this transformation $\omega'=W+\Delta\omega/2\,,\, \omega=W-\Delta\omega/2$,
\begin{equation}
    P_{\eta}(W, \Delta\omega, X) = \gamma^2 \xi^2\left(W+\frac{\Delta\omega}{2}\right) \xi^2\left(W-\frac{\Delta\omega}{2}\right)
    \frac{ 1 + \alpha(X) \eta^2 \cos\left(\Delta\omega \Delta t\right)}{2}.
\end{equation}
It is important to notice that the chosen transformation has an unitary Jacobian, and therefore the probability do not need to be re-normalized. Also, it is possible to factorize the probabilities in two parts: the first is a function of $\Delta\omega$ and the second is a function of $W$. In fact, we can write
\begin{align}
\begin{split}
    \xi^2\left(W+\frac{\Delta\omega}{2}\right) \xi^2\left(W-\frac{\Delta\omega}{2}\right)&=\frac{1}{2\pi\sigma^2}\exp\left[-\frac{(W-\Omega_0+\Delta\omega/2)^2}{2\sigma^2}\right]\exp\left[-\frac{(W-\Omega_0-\Delta\omega/2)^2}{2\sigma^2}\right]\\
    &=\frac{1}{2\pi\sigma^2}\exp\left[-\frac{(W-\Omega_0)^2}{\sigma^2}\right]\exp\left[-\frac{\Delta\omega^2}{4\sigma^2}\right],
    \end{split}
\end{align}
and therefore we obtain
\begin{align}
\begin{split}
     P_{\eta}(W, \Delta\omega, X)& =  \frac{\gamma^2}{2\pi\sigma^2}\exp\left[-\frac{(W-\Omega_0)^2}{\sigma^2}\right]\exp\left[-\frac{\Delta\omega^2}{4\sigma^2}\right]
   \frac{ 1 + \alpha(X) \eta^2 \cos\left(\Delta\omega \Delta t\right)}{2}\\
   &=G_{\Omega_0,\sigma^2}(W)P_{\eta}(\Delta\omega, X),
   \end{split}
\end{align}
where $G_{\Omega_0,\sigma^2}(W)$ in the Gaussian distribution for the variable $W$ with parameters $\Omega_0,\sigma^2$, and
\begin{equation}
   P_{\eta}(\Delta\omega, X) =\int dW P_{\eta}(W, \Delta\omega, X)=\gamma^2C(\Delta\omega)\frac{ 1 + \alpha(X) \eta^2 \cos\left(\Delta\omega \Delta t\right)}{2},
\end{equation}
where 
\begin{equation}
    C(\Delta\omega)=\frac{1}{\sqrt{4\pi\sigma^2}}\exp\left[-\frac{\Delta\omega^2}{4\sigma^2}\right].
\end{equation}
It is possible to show that measuring $W$ does not improve the amount of information for the estimation of $\Delta t$ by proving that the Fisher information evaluated by using $P_{\eta}(\Delta\omega, X)$ is equal to the one evaluated by using and $P_{\eta}(W, \Delta\omega, X)$,
\begin{align}
\begin{split}
    \sum_{X=B,C}\int dWd\omega \frac{(\partial_{\Delta t}P_{\eta}(W,\omega, X))^2}{P_{\eta}(W,\omega, X)}&=\sum_{X=B,C}\int dWd\omega \frac{(G_{\Omega_0,\sigma^2}(W)\partial_{\Delta t}P_{\eta}(\omega, X))^2}{G_{\Omega_0,\sigma^2}(W)P_{\eta}(\omega, X)}\\
    &=\sum_{X=B,C}\int dWG_{\Omega_0,\sigma^2}(W)\int d\omega \frac{(\partial_{\Delta t}P_{\eta}(\omega, X))^2}{P_{\eta}(\omega, X)}\\
    &=\sum_{X=B,C}\int d\omega \frac{(\partial_{\Delta t}P_{\eta}(\omega, X))^2}{P_{\eta}(\omega, X)}.
    \end{split}
\end{align}
\section{Finite resolution probabilities} \label{SM_finite_freq}
In this section, we will study the case in which the detectors have a finite resolution $\epsilon$, such as the detector will associate the frequency shift between the two photons to bins of width $2\epsilon$. the bins are ordered from $0$ to $n_{\mathrm{MAX}}$, where $(n_{\mathrm{MAX}}+1)\epsilon$ is the maximum range for the detection of the frequency shift of the two photons.   
For simplicity, we redefine the probability distribution with infinite resolution
\begin{equation}
\begin{aligned}
P\bigl(X,\tau,\omega,\sigma,\eta\bigr)
= \frac{1}{\sqrt{4\pi\sigma^2}}
\exp\!\biggl(-\frac{\omega^2}{4\sigma^2}\biggr)\,
\frac{1 + \alpha(X)\,\eta\,\cos(\tau\,\omega)}{2}\,.
\end{aligned}
\end{equation}
When we consider a finite resolution of the detectors, we need to integrate the probabilities with a kernel that express the response of the detector. We model this response with a triangular function with the peak at the center of the beam. For the first bin, i.e., the one peaked in 0, we  have the probability
\begin{equation}
\begin{aligned}
\begin{split}
P_{0}\bigl(X,\tau,\epsilon,\sigma,\eta\bigr)
&=\int_{0}^{\epsilon}
   \Bigl(1 - \tfrac{\omega}{\epsilon}\Bigr)\,
   P\bigl(X,\tau,\omega,\sigma,\eta\bigr)\,\mathrm{d}\omega\\
   &=\frac{1}{\sqrt{\pi}}\int_0^{\epsilon'}\Bigl(1 - \tfrac{\varpi}{\epsilon'}\Bigr)\mathrm{e}^{-\varpi^2}\frac{1+\alpha(X)\eta\cos(2\sigma\tau\varpi)}{2}d\varpi,
   \end{split}
\end{aligned}
\end{equation}
where $\epsilon'=\epsilon/2\sigma$, and $\varpi=\omega/2\sigma$, while, for the $n$-th bin, $n>0$, we have
\begin{equation}
\begin{aligned}
\begin{split}
P_{n>0}\bigl(X,\tau,\epsilon,\sigma,\eta\bigr)
&= \int_{(n-1)\epsilon}^{(n+1)\epsilon}
   f_n\left(\frac{\omega}{\epsilon}\right)\,
   P\bigl(X,\tau,\omega,\sigma,\eta\bigr)\,\mathrm{d}\omega\\
   &=\frac{1}{\sqrt{\pi}}\int_{(n-1)\epsilon'}^{(n+1)\epsilon'}
   f_n\left(\frac{\varpi}{\epsilon'}\right)\,
   \mathrm{e}^{-\varpi^2}\frac{1+\alpha(X)\eta\cos(2\sigma\tau\varpi)}{2}d\varpi
   \end{split}
\end{aligned}
\end{equation}
where
\begin{equation}
    f_n\left(\frac{\omega}{\epsilon}\right)=\begin{cases}
        \frac{\omega}{\epsilon}-n+1\qquad \mathrm{for}\qquad (n-1)\epsilon\leq\omega< n\epsilon\\
        n+1-\frac{\omega}{\epsilon}\qquad \mathrm{for}\qquad n\epsilon\leq\omega\leq (n+1)\epsilon
    \end{cases}\qquad.
\end{equation}
The Fisher information for the parameter $\tau$, $F(\tau)$, will be the sum of all the $n$-th contribute $F_n(\tau)$. Each of these contributions is associated with the $n$-th bin of the new probability set. Therefore we have
\begin{equation}
    F(\tau)=\sum_{n=0}^{n_{\mathrm{MAX}}}F_n(\tau),
    \label{SM_FI_tot}
\end{equation}
where
\begin{equation}
    F_n(\tau)=\sum_{X=A,B}\frac{1}{P_{0}\bigl(X,\tau,\epsilon,\sigma,\eta\bigr)}\left(\frac{\partial P_{0}\bigl(X,\tau,\epsilon,\sigma,\eta\bigr)}{\partial\tau}\right)^2,
\end{equation} \label{SM_FI_zero}

Starting with the evaluation of $F_0(\tau)$, we have
\begin{align}
    \begin{split}
        F_0(\tau)&=\sum_{X=A,B}\frac{1}{P_{0}\bigl(X,\tau,\epsilon,\sigma,\eta\bigr)}\left(\frac{\partial P_{0}\bigl(X,\tau,\epsilon,\sigma,\eta\bigr)}{\partial\tau}\right)^2\\
        &=\sum_{X=A,B}\frac{1}{\int_{0}^{\epsilon}
   \Bigl(1 - \tfrac{\omega}{\epsilon}\Bigr)\,
   P\bigl(X,\tau,\omega,\sigma,\eta\bigr)\,\mathrm{d}\omega}\left(\frac{\partial }{\partial\tau}\int_{0}^{\epsilon}
   \Bigl(1 - \tfrac{\omega}{\epsilon}\Bigr)\,
   P\bigl(X,\tau,\omega,\sigma,\eta\bigr)\,\mathrm{d}\omega\right)^2
    \end{split}
\end{align}
We notice that the numerator of each term of the sum does not depend on $X$, and therefore can be factorized. In fact
\begin{equation}
\begin{split}
    \left(\frac{\partial }{\partial\tau}\int_{0}^{\epsilon}
   \Bigl(1 - \tfrac{\omega}{\epsilon}\Bigr)\,
   P\bigl(X,\tau,\omega,\sigma,\eta\bigr)\,\mathrm{d}\omega\right)^2&=\left(\frac{\sigma\eta}{\sqrt{\pi}}\int_{0}^{\epsilon'}
   \Bigl(1 - \tfrac{\varpi}{\epsilon'}\Bigr)\,
   \mathrm{e}^{-\varpi^2}\varpi\sin(2\sigma\tau\varpi)\,\mathrm{d}\varpi\right)^2.
   \end{split}
\end{equation}
So, we will obtain
\begin{equation}
\begin{split}
         F_0(\tau)&=\left(\frac{\sigma\eta}{\sqrt{\pi}}\int_{0}^{\epsilon'}
   \Bigl(1 - \tfrac{\varpi}{\epsilon'}\Bigr)\,
   \mathrm{e}^{-\varpi^2}\varpi\sin(2\sigma\tau\varpi)\,\mathrm{d}\varpi\right)^2\sum_{X=A,B}\frac{1}{\int_{0}^{\epsilon}
   \Bigl(1 - \tfrac{\omega}{\epsilon}\Bigr)\,
   P\bigl(X,\tau,\omega,\sigma,\eta\bigr)\,\mathrm{d}\omega}\\
   &=\frac{\left(\frac{\sigma\eta}{\sqrt{\pi}}\int_{0}^{\epsilon'}
   \Bigl(1 - \tfrac{\varpi}{\epsilon'}\Bigr)\,
   \mathrm{e}^{-\varpi^2}\varpi\sin(2\sigma\tau\varpi)\,\mathrm{d}\varpi\right)^2\left(\frac{1}{\sqrt{\pi}}\int_{0}^{\epsilon'}
   \Bigl(1 - \tfrac{\varpi}{\epsilon'}\Bigr)\,
   \mathrm{e}^{-\varpi^2}d\varpi\right)}{\left(\frac{1}{2\sqrt{\pi}}\int_{0}^{\epsilon'}
   \Bigl(1 - \tfrac{\varpi}{\epsilon'}\Bigr)\,
   \mathrm{e}^{-\varpi^2}d\varpi\right)^2-\left(\frac{\eta}{2\sqrt{\pi}}\int_{0}^{\epsilon'}
   \Bigl(1 - \tfrac{\varpi}{\epsilon'}\Bigr)\,
   \mathrm{e}^{-\varpi^2}\cos(2\sigma\tau\varpi)\,\mathrm{d}\varpi\right)^2}\\
   &=\frac{4\sigma^2\eta^2}{\sqrt{\pi}}\frac{\left(\int_{0}^{\epsilon'}
   \Bigl(1 - \tfrac{\varpi}{\epsilon'}\Bigr)\,
   \mathrm{e}^{-\varpi^2}\varpi\sin(2\sigma\tau\varpi)\,\mathrm{d}\varpi\right)^2\left(\int_{0}^{\epsilon'}
   \Bigl(1 - \tfrac{\varpi}{\epsilon'}\Bigr)\,
   \mathrm{e}^{-\varpi^2}d\varpi\right)}{\left(\int_{0}^{\epsilon'}
   \Bigl(1 - \tfrac{\varpi}{\epsilon'}\Bigr)\,
   \mathrm{e}^{-\varpi^2}d\varpi\right)^2-\left(\eta\int_{0}^{\epsilon'}
   \Bigl(1 - \tfrac{\varpi}{\epsilon'}\Bigr)\,
   \mathrm{e}^{-\varpi^2}\cos(2\sigma\tau\varpi)\,\mathrm{d}\varpi\right)^2}.
\end{split}
\end{equation}
To summarize, we can write $F_0(\tau)$ as follows
\begin{equation}   F_0(\tau)=H\frac{2\eta^2I_{0,\mathrm{num}}}{\sqrt{\pi}I_{0,\mathrm{den}}},
\end{equation} \label{Fin_FI_less}
where $H=2\sigma^2$ is the quantum Fisher information,
\begin{equation}
    I_{0,\mathrm{num}}=\left(\int_{0}^{\epsilon'}
   \Bigl(1 - \tfrac{\varpi}{\epsilon'}\Bigr)\,
   \mathrm{e}^{-\varpi^2}\varpi\sin(2\sigma\tau\varpi)\,\mathrm{d}\varpi\right)^2\left(\int_{0}^{\epsilon'}
   \Bigl(1 - \tfrac{\varpi}{\epsilon'}\Bigr)\,
   \mathrm{e}^{-\varpi^2}d\varpi\right)
\end{equation}
and
\begin{equation}
  I_{0,\mathrm{den}}= \left(\int_{0}^{\epsilon'}
   \Bigl(1 - \tfrac{\varpi}{\epsilon'}\Bigr)\,
   \mathrm{e}^{-\varpi^2}d\varpi\right)^2-\left(\eta\int_{0}^{\epsilon'}
   \Bigl(1 - \tfrac{\varpi}{\epsilon'}\Bigr)\,
   \mathrm{e}^{-\varpi^2}\cos(2\sigma\tau\varpi)\,\mathrm{d}\varpi\right)^2 .
\end{equation}
Analogously, we can find $F_{n>0}(\tau)$ in the form
\begin{equation}
    F_{n>0}(\tau)=H\frac{2\eta^2I_{n>0,\mathrm{num}}}{\sqrt{\pi}I_{n>0,\mathrm{den}}},
\end{equation} \label{Fin_FI_great}
where
\begin{equation}
    I_{n>0,\mathrm{num}}=\left(\int_{(n-1)\epsilon'}^{(n+1)\epsilon'}
   f_n\left(\frac{\varpi}{\epsilon'}\right)\,
   \mathrm{e}^{-\varpi^2}\varpi\sin(2\sigma\tau\varpi)\,\mathrm{d}\varpi\right)^2\left(\int_{(n-1)\epsilon'}^{(n+1)\epsilon'}
   f_n\left(\frac{\varpi}{\epsilon'}\right)\,
   \mathrm{e}^{-\varpi^2}d\varpi\right)
\end{equation}
and
\begin{equation}
  I_{n>0,\mathrm{den}}= \left(\int_{(n-1)\epsilon'}^{(n+1)\epsilon'}
   f_n\left(\frac{\varpi}{\epsilon'}\right)\,
   \mathrm{e}^{-\varpi^2}d\varpi\right)^2-\left(\eta\int_{(n-1)\epsilon'}^{(n+1)\epsilon'}
   f_n\left(\frac{\varpi}{\epsilon'}\right)\,
   \mathrm{e}^{-\varpi^2}\cos(2\sigma\tau\varpi)\,\mathrm{d}\varpi\right)^2 .
\end{equation}

\end{document}